\documentclass[10pt, conference, compsocconf]{IEEEtran}

\usepackage{cite}

\usepackage{graphicx}

\usepackage[cmex10]{amsmath}

\usepackage{listings}
\usepackage{array}

\usepackage{dblfloatfix}

\usepackage{tikz}
\usetikzlibrary{positioning,shapes,external}

\tikzset{external/prefix={cache/}}
\tikzset{external/mode=list and make}

\definecolor{kit-green100}{rgb}{0,.59,.51}
\definecolor{kit-blue100}{rgb}{.27,.39,.67}
\definecolor{kit-red}{RGB}{160,30,40}

\colorlet{dblue}{kit-blue100!60!blue}
\colorlet{dgreen}{kit-green100}
\colorlet{dred}{kit-red!60!red}
\colorlet{dpurple}{violet}

\usepackage{pgfplots}

\pgfplotscreateplotcyclelist{color}{%
  { blue, thick, solid, mark=+, every mark/.append style={scale=0.95} },
  { red, thick, solid, mark=x, every mark/.append style={scale=0.95} },
  { green!70!black, thick, solid, mark=triangle, every mark/.append style={scale=0.8} },
  { purple, thick, solid, mark=square, every mark/.append style={scale=0.7} },
  { orange, thick, solid, mark=o, every mark/.append style={scale=0.7} },
  { cyan, thick, solid, mark=square*, every mark/.append style={scale=0.8} },
}

\definecolor{c1}{RGB}{255,0,0}
\definecolor{c2}{HTML}{22A8F0}
\definecolor{c3}{HTML}{99D0A5}

\pgfplotscreateplotcyclelist{color_nomark}{%
  { c1, thick, solid  },
  { c2, thick, solid },
  { c3, thick },
  { purple, thick, solid },
  { orange, thick, solid },
  { cyan, thick, solid },
}

\pgfplotsset{
  myPlot/.style={
    grid,
    width=63mm,height=51mm,
    major grid style={thin,dotted,color=black!50},
    minor grid style={thin,dotted,color=black!50},
    every axis/.append style={
      thin,
      tick style={
        line cap=round,
        thin,
      },
    },
    cycle list name={color},
    major tick length=3pt,
    minor tick length=1.5pt,
    legend cell align=left,
    ylabel near ticks,
    enlarge x limits=0.08,
    xtick={1,2,4,8,16},
    x tick label style={/pgf/number format/.cd, set thousands separator={\,}},
    y tick label style={/pgf/number format/.cd, set thousands separator={\,}},
  },
  myPlotProfile/.style={
    myPlot,
    width=166mm,height=41mm,
    xtick={},
    enlarge x limits=0,
    cycle list name={color_nomark},
    title style={yshift=-1.0ex},
  },
  myPlotProfileLeft/.style={
    myPlotProfile,
    axis y line*=left,
    ylabel={CPU Utilization [\%]},
    ymin=0, ymax=100, grid=none,
  },
  myPlotProfileRight/.style={
    myPlotProfile,
    axis y line*=right,
    axis x line=none,
    ylabel={Network/Disk\\ Throughput [MiB/s]},
    ylabel style={rotate=180,align=center},
    ymin=0,
  },
}

\usepackage{url}

\def\textcode#1{\emph{#1}}
\def\Arr#1{[{#1}]}
\def\DIA#1{\textcode{DIA$\langle${#1}$\rangle$}}
\def\DIAOp#1{\textcode{#1}}

\begin{document}

\title{Thrill: High-Performance Algorithmic Distributed Batch Data Processing with C++}

\author{\IEEEauthorblockN{%
    Timo Bingmann\IEEEauthorrefmark{1},
    Michael Axtmann\IEEEauthorrefmark{1},
    Emanuel J\"obstl,
    Sebastian Lamm,
    Huyen Chau Nguyen,\\
    Alexander Noe,
    Sebastian Schlag\IEEEauthorrefmark{1},
    Matthias Stumpp, 
    Tobias Sturm, and
    Peter Sanders\IEEEauthorrefmark{1}
  }
  \IEEEauthorblockA{%
    Institute of Theoretical Informatics\\
    Karlsruhe Institute of Technology\\
    Karlsruhe, Germany\\
    \IEEEauthorrefmark{1}Emails: \{firstname.lastname\}@kit.edu
  }
}


\maketitle

\begin{abstract}
  We present the design and a first performance evaluation of Thrill -- a prototype of a general purpose big data processing framework with a convenient data-flow style programming interface.
  Thrill is somewhat similar to Apache Spark and Apache Flink with at least two main differences.
  First, Thrill is based on C++ which enables performance advantages due to direct native code compilation, a more cache-friendly memory layout, and explicit memory management.
  In particular, Thrill uses template meta-programming to compile chains of subsequent local operations into a single binary routine without intermediate buffering and with minimal indirections.
  Second, Thrill uses arrays rather than multisets as its primary data structure which enables additional operations like sorting, prefix sums, window scans, or combining corresponding fields of several arrays (zipping).

  We compare Thrill with Apache Spark and Apache Flink using five kernels from the HiBench suite.
  Thrill is consistently faster and often several times faster than the other frameworks.
  At the same time, the source codes have a similar level of simplicity and abstraction.
\end{abstract}



\section{Introduction}

In this paper we present Thrill, a new open-source C++ framework for algorithmic distributed batch data processing.

The need for parallel and distributed algorithms cannot be ignored anymore, since individual processor cores' clock speeds have stagnated in recent years.
At the same time, we have experienced an explosion in data volume so that scalable distributed data analysis has become a bottleneck in an ever-increasing range of applications.
With Thrill we want to make a step at bridging the gap between two traditional scenarios of ``Big Data'' processing.

On the one hand, in academia and high-performance computing (HPC), distributed algorithms are often handcrafted in C/C++ and use MPI for explicit communication. This achieves high efficiency at the price of difficult implementation.
On the other hand, global players in the software industry created their own ecosystem to cope with their data analysis needs.
Google popularized the MapReduce~\cite{dean2008mapreduce} model in 2004 and described their in-house implementation.
Apache Hadoop and more recently Apache Spark~\cite{zaharia2010spark} and Apache Flink~\cite{alexandrov2014stratosphere} have gained attention as open-source Scala/Java-based solutions for heavy duty data processing.
These frameworks provide a simple programming interface and promise \emph{automatic} work parallelization and scheduling, \emph{automatic} data distribution, and \emph{automatic} fault tolerance.
While most benchmarks highlight the scalability of these frameworks, the bottom line efficiency has been shown to be lacking~\cite{mcsherry2015scalability}, surprisingly with the CPU often being the bottleneck~\cite{ousterhout2015making}.

Thrill's approach to bridging this gap is a library of \emph{scalable algorithmic primitives} like \DIAOp{Map}, \DIAOp{ReduceByKey}, \DIAOp{Sort}, and \DIAOp{Window}, which can be combined efficiently to construct larger complex algorithms using pipelined data-flow style programming.
Thrill is written in modern C++14 from the ground up, has minimal external dependencies, and compiles cross-platform on Linux, Mac OS, and Windows.
By using C++, Thrill is able to exploit compile-time optimization, template meta-programming, and explicit memory management.
Thrill enables efficient processing of fixed-length items like single characters or fixed-dimensional vectors without object overhead due to the zero overhead abstractions of C++.
It treats data types of operations as opaque and utilizes template programming to instantiate operations with user-defined functions (UDFs).
For example, the comparison function of the sorting operation is compiled into the actual internal sorting and merging algorithms (similar to \textcode{std::sort}).
At the same time, Thrill makes no attempts to optimize the execution order of operations, as this would require introspection into the data and how UDFs manipulate it.

Thrill programs run in a collective bulk-synchronous manner similar to most MPI programs.
Thrill primarily focuses on fast in-memory computation, but transparently uses external memory when needed.
The functional programming style used by Thrill enables easy parallelization, which also works remarkably well for shared memory parallelism.
Hence, due to the restriction to scalable primitives, Thrill programs run on a wide range of homogeneous parallel systems.

By using C++, Thrill aims for high performance distributed algorithms. JVM-based frameworks are often slow due to the overhead of the interpreted bytecode, even though just-in-time (JIT) compilation has leveled the field somewhat. Nevertheless, due to object indirections and garbage collection, Java/Scala must remain less cache-efficient.
While efficient CPU usage should be a matter of course, especially when processing massive amounts of data, the ultimate bottleneck for scalable distributed application is the (bisection) bandwidth of the network.
But by using more tuned implementations, more CPU time is left for compression, deduplication~\cite{sanders2013communication}, and other algorithms to reduce communication.
Nevertheless, in smaller networks the CPU is often the bottleneck~\cite{ousterhout2015making}, and for most applications a small cluster is sufficient.

A consequence of using C++ is that memory management has to be done explicitly. While this is desirable for more predictable and higher performance than garbage collected memory, it does make programming more difficult. However, with modern C++11 this has been considerably alleviated, and Thrill uses reference counting extensively.

While scalable algorithms promise eventually higher performance with more hardware, the performance hit going from parallel shared memory to a distributed setting is large.
This is due to the communication latency and bandwidth bottlenecks.
This network overhead and the additional management overheads of big data frameworks often make speedups attainable only with unjustifiable hardware costs~\cite{mcsherry2015scalability}.
Thrill cannot claim zero overhead, as network costs are unavoidable. But by overlapping computation and communication, and by employing binary optimized machine code, we keep the overhead small.

Thrill is open-source under the BSD 2-clause license and available as a community project on GitHub\footnote{\url{http://github.com/thrill/thrill}}. It currently has more than 52\,K lines of C++ code and approximately a dozen developers have contributed.

\paragraph*{Overview} The rest of this section introduces related work with an emphasis on Spark and Flink.
Section~\ref{sec:design} discusses the design of Thrill, in particular its API and the rationale behind the chosen concept.
We present a complete WordCount example in Section~\ref{sec:wordcount}, followed by an overview of the current portfolio of operations and details of their implementation.
In Section~\ref{sec:experiments}, experimental results of a comparison of Thrill, Spark, and Flink based on five micro benchmarks including PageRank and KMeans are shown.
Section~\ref{sec:conclusion} concludes and provides an outlook for future work.

\paragraph*{Our Contributions}
Thrill demonstrates that with the advent of C++11 lambda-expressions, it has become feasible to use C++ for big data processing using an abstract and convenient API comparable to currently popular frameworks like Spark or Flink.
This not only harvests the usual performance advantages of C++, but allows us moreover to transparently compile sequences of local operations into a single binary code via sophisticated template meta-programming.
By using arrays as the primary data type, we enable additional basic operations that have to be emulated by more complicated and more costly operations in traditional multiset-based systems.
Our experimental evaluation demonstrates that even the current prototypical implementation already offers a considerable performance advantage over Spark and Flink.

\subsection{Related Work}

Due to the importance and hype of the ``Big Data'' topic, a myriad of distributed data processing frameworks have been proposed in recent years~\cite{chen2014data}.
These cover many different aspects of this challenge like data warehousing and batch processing, stream aggregation~\cite{toshniwal2014storm}, interactive queries~\cite{melnik2010dremel}, and specialized graph~\cite{malewicz2010pregel,low2012distributed} and machine learning frameworks~\cite{tensorflow2015-whitepaper}.

In 2004, Google established the MapReduce paradigm~\cite{dean2008mapreduce} as an easy-to-use interface for scalable data analysis.
Their paper spawned a whole research area on how to express distributed algorithms using just \emph{map} and \emph{reduce} in as few rounds as possible.
Soon, Apache Hadoop was created as an open-source MapReduce framework written in Java for commodity hardware clusters.
Most notable from this collection of programs was the Hadoop distributed file system (HDFS)~\cite{shvachko2010hadoop}, which is key for fault tolerant data management for MapReduce.
Subsequently, a large body of academic work was done optimizing various aspects of Hadoop like scheduling and data shuffling~\cite{lee2012parallel}.

MapReduce and Hadoop are very successful due to their simple programming interface, which at the same time is a severe limitation.
For example, iterative computations are reported to be very slow due to the high number of MapReduce rounds, each of which may need a complete data exchange and round-trip to disks.
More recent frameworks such as Apache Spark and Apache Flink offer a more general interface to increase usability and performance.


Apache Spark operates on an abstraction called \emph{resilient distributed datasets} (RDDs)~\cite{zaharia2010spark}.
This abstraction provides the user with an easy-to-use interface which consists of a number of deterministic coarse-grained operations.
Each operation can be classified either as \emph{transformation} or \emph{action}.
A transformation is a lazy operation that defines a new RDD given another one, e.g. \DIAOp{map} or \DIAOp{join}.
An action returns computed results to the user program, e.g. \DIAOp{count}, \DIAOp{collect}, or reads/writes data from/to external storage.
When an action triggers computation, Spark examines the sequence of previously called transformations and identifies so-called execution \emph{stages}.
Spark runs in a master-worker architecture. While the driver program runs on the master, the actual computation occurs on the workers with a \emph{block-based} work-partitioning and scheduling system.
Spark can maintain already computed RDDs in main memory to be reusable by future operations, in order to speed-up iterative computations~\cite{zaharia2012resilient}.

In more recent versions, Spark added two more APIs: DataFrames~\cite{armbrust2015spark} and Datasets.
Both offer domain specific languages for higher level declarative programming similar to SQL, which allows Spark to optimize the query execution plan.
Even further, it enables Spark to generate optimized query bytecode online, aside of the original Scala/Java program.
The optimized bytecode can use more efficient direct access methods to the data, which no longer needs to be stored as JVM objects, and hence garbage collection can be avoided.
The DataFrame engine is built on top of the original RDD processing interface.


Apache Flink originated from the Stratosphere research project~\cite{alexandrov2014stratosphere} and is progressing from an academic project to industry.
While Flink shares many ideas with Spark such as the master-worker model, lazy operations, and iterative computations, it tightly integrates concepts known from parallel database systems.
Flink's core interface is a domain-specific declarative language.
Furthermore, Flink's focus has turned to streaming rather than batch processing.

In Flink, the optimizer takes a user program and produces a graph of logical operators.
The framework then performs rule- and cost-based optimizations, such as reordering of operations, pipelining of local operations, selection of algorithms, and evaluation of different data exchange patterns to find an execution plan Flink believes is best for a given user program and cluster configuration.
Flink is based on a pipelined execution engine comparable to parallel database systems, which is extended to integrate streaming operations with rich windowing semantics.
Iterative computations are sped up by performing delta-iterations on changing data only, as well as placing computation on the same worker across iterations.
Fault tolerance is achieved by continuously taking snapshots of the distributed data streams and operator states~\cite{carbone2015lightweight}, a concept inspired by Chandy-Lamport snapshots~\cite{chandy1985distributed}.
Flink also has  an own memory management system separate from the JVM's garbage collector for higher performance and better control of memory peaks.


The interfaces of Spark and Flink differ in some very important ways.
Flink's optimizer requires introspection into the components of data objects and how the UDFs operate on them.
This requires many Scala/Java annotations to the UDFs and incurs an indirection for access to the values of components.
In contrast to Spark's RDD interface, where users can make use of \emph{host language control-flow}, Flink provides custom iteration operations.
Hence, Flink programs are in this respect more similar to declarative SQL statements than to an imperative language.
The newer DataFrame and Dataset interfaces introduce similar concepts to Spark, but extend them further with a custom code generation engine.
At its core, Spark is an in-memory batch engine that executes streaming jobs as a series of mini-batches.
In contrast, Flink is based on a pipelined execution engine used in database systems, allowing Flink to process streaming operations in a pipelined way with lower latency than in the micro-batch model.
In addition, Flink supports external memory algorithms whereas Spark is mainly an in-memory system with spilling to external memory.


Overall the JVM is currently the dominant platform for open-source big data frameworks.
This is understandable from the point of view of programmer productivity but surprising when considering that C++ is the predominant language for performance critical systems -- and big data processing is inherently performance critical.
Spark~\cite{armbrust2015scaling} (with Project Tungsten) and Flink (with MemorySegments) therefore put great efforts into overcoming performance penalties of the JVM, for example by using explicit \emph{Unsafe} memory operations and generating optimized bytecode to avoid object overhead and garbage collection.
With Thrill we present a C++ framework as an alternative that does not incur these overheads in the first place.


\section{Design of Thrill}\label{sec:design}

Thrill programs are written in C++ and compile into binary programs.
The execution model of this binary code is similar to MPI programs: one identical program is run collectively on $h$ machines.
Thrill currently expects all machines to have nearly identical hardware, since it balances work and data equally between the machines.
The binary program is started simultaneously on all machines, and connects to the others via a network protocol.
Thrill currently supports TCP sockets and MPI as network backends.
The startup procedures depend on the specific backend and cluster environment.

Each machine is called a \emph{host}, and each work thread on a host is called a \emph{worker}. Currently, Thrill requires all hosts to have the same number of cores $c$, hence, in total there are $p = h \cdot c$ worker threads. Additionally, each host has one thread for network/data handling and one for asynchronous disk I/O.
Each of the $h$ hosts have $h-1$ reliable network connections to the other hosts, and the hosts and workers are enumerated $0 \ldots h-1$ and $0 \ldots p-1$. Thrill does not have a designated master or driver host, as all communication and computation is done collectively.

Thrill currently provides no fault tolerance. While our data-flow API permits smooth integration of fault tolerance using asynchronous checkpoints~\cite{carbone2015lightweight,chandy1985distributed}, the execution model of exactly $h$ machines may have to be changed.

\subsection{Distributed Immutable Arrays}

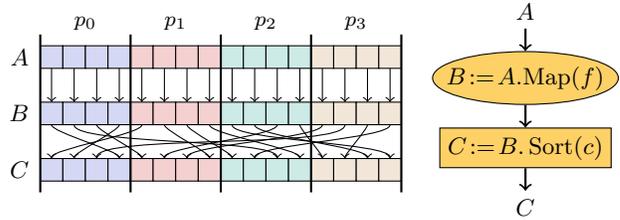
\begin{figure}\centering\small
  \begin{tikzpicture}[scale=0.3,
    item/.style={draw, minimum size=10mm, transform shape},
    itemPE0/.style={item, fill=dblue!20},
    itemPE1/.style={item, fill=dred!20},
    itemPE2/.style={item, fill=dgreen!20},
    itemPE3/.style={item, fill=brown!20},
    item0/.style={itemPE0}, item1/.style={itemPE0}, item2/.style={itemPE0}, item3/.style={itemPE0},
    item4/.style={itemPE1}, item5/.style={itemPE1}, item6/.style={itemPE1}, item7/.style={itemPE1},
    item8/.style={itemPE2}, item9/.style={itemPE2}, item10/.style={itemPE2}, item11/.style={itemPE2},
    item12/.style={itemPE3}, item13/.style={itemPE3}, item14/.style={itemPE3}, item15/.style={itemPE3},
    ]

    \foreach \x in {0,...,15} {
      \node[item\x] (A\x) at (\x,0) {};
    }

    \node[left=1pt of A0] {$A$};

    \begin{scope}[yshift=-25mm]
      \foreach \x in {0,...,15} {
        \node[item\x] (B\x) at (\x,0) {};
      }
    \end{scope}

    \foreach \x in {0,...,15} {
      \draw[->] (A\x) -- (B\x);
    }

    \node[left=1pt of B0] {$B$};

    \begin{scope}[yshift=-5cm]
      \foreach \x in {0,...,15} {
        \node[item\x] (C\x) at (\x,0) {};
      }
    \end{scope}

    \node[left=1pt of C0] {$C$};

    \foreach \x in {0,4,8,12,16} {
      \draw[thick] (\x - 0.5,1) -- (\x - 0.5,-6);
    }

    \begin{scope}[font=\footnotesize,yshift=15mm]
      \node at (1.5,0) {$p_0$};
      \node at (5.5,0) {$p_1$};
      \node at (9.5,0) {$p_2$};
      \node at (13.5,0) {$p_3$};
    \end{scope}

    \foreach \x/\y in {%
      0/3, 1/10, 2/4, 3/1, 4/0, 5/7, 6/8, 7/11, 8/15,
      9/14, 10/5, 11/12, 12/6, 13/2, 14/13, 15/9}
    {
      \draw[->] (B\x.south) to[out=270,in=90,looseness=0.25] (C\y.north);
    }

    \begin{scope}[
      xshift=210mm, yshift=20mm,
      node distance=3mm,
      datanode/.style={draw, fill=yellow!50!orange!70!white},
      dataarrow/.style={->, thick},
      ]

      \node (A) {$A$};
      \node [datanode, ellipse, below=of A, inner sep=-5pt, minimum height=20pt] (B) {$B \!:=\! A.\!\operatorname{Map}(f)$};
      \node [datanode, rectangle, below=of B, inner sep=3pt] (C) {$C \!:=\! B.\operatorname{Sort}(c)$};
      \node [below=of C] (C') {$C$};

      \draw[dataarrow] (A) -- (B);
      \draw[dataarrow] (B) -- (C);
      \draw[dataarrow] (C) -- (C');

    \end{scope}
  \end{tikzpicture}
  \vspace*{-3mm}
  \caption{Distribution of a DIA between processors (left) and a data-flow graph (right)}
  \label{alg:dia graph}
\end{figure}

The central concept in Thrill's high-level data-flow API is the \emph{distributed immutable array} (DIA).
A DIA is an array of items which is distributed over the cluster in some way.
No direct array access is permitted.
Instead, the programmer can apply so-called \emph{DIA operations} to the array as a whole.
These operations are a set of \emph{scalable primitives}, listed in Table~\ref{tab:operations}, which can be composed into complex distributed algorithms.
DIA operations can create DIAs by reading files, transform existing DIAs by applying user functions, or calculate scalar values collectively, used to determine the further program control flow. In a Thrill program, these operations are used to lazily construct a DIA \emph{data-flow} graph in C++ (see Figure~\ref{alg:dia graph}).
The data-flow graph is only executed when an \emph{action} operation is encountered.
How DIA items are actually stored and in what way the operations are executed on the distributed system remains transparent to the user.

In the current prototype of Thrill, the array is usually distributed evenly between the $p$ workers in order. DIAs can contain any C++ data type, provided serialization methods are available (more in Section~\ref{sec:data}). Thrill contains built-in serialization methods for all primitive types, and most STL types; only custom non-trivial classes require additional methods.
Each DIA operation in Table~\ref{tab:operations} is implemented as a C++ template class, which can be instantiated with appropriate UDFs.

\subsection{Example: WordCount}\label{sec:wordcount}

\begin{figure}\centering
\begin{lstlisting}[numbers=left, numberstyle=\tiny, numbersep=4pt]
void WordCount(thrill::Context& ctx,
  std::string input, std::string output) {
  using Pair = std::pair<std::string, size_t>;
  auto word_pairs = ReadLines(ctx, input)
  .template FlatMap<Pair>(
    // flatmap lambda: split and emit each word
    [](const std::string& line, auto emit) {
      Split(line, ' ', [&](std::string_view sv) {
        emit(Pair(sv.to_string(), 1));
      });
    });
  word_pairs.ReduceByKey(
    // key extractor: the word string
    [](const Pair& p) { return p.first; },
    // commutative reduction: add counters
    [](const Pair& a, const Pair& b) {
      return Pair(a.first, a.second + b.second);
    })
  .Map([](const Pair& p) {
    return p.first + ": "
           + std::to_string(p.second); })
  .WriteLines(output);
}
\end{lstlisting}
\vspace*{-3mm}
\caption{Complete WordCount Example in Thrill}
\label{alg:wc}
\end{figure}

We now present a complete code of the popular WordCount benchmark in Algorithm~\ref{alg:wc} to demonstrate how easy it is to program in Thrill. The program counts the number of occurrences of each unique word in a text.
In Thrill, WordCount including file I/O consists of five DIA operations.

\DIAOp{ReadLines} (line~4) and \DIAOp{WriteLines} (line~22) are used to read the text and write the result from/to the file system.
Thrill currently uses standard POSIX filesystem methods to read and write to disk, and it requires a distributed parallel file system such as NFS, Lustre, or Ceph to provide a common view to all compute hosts.
ReadLines takes a \textcode{thrill::Context} object, which is only required for source DIA operations, and a set of files.
The result of ReadLines is a \DIA{std::string}, which contains each line of the files as an item.
The set of files is ordered lexicographically and the set of lines is partitioned equally among the workers.

However, this DIA is not assigned to a variable name.
Instead, we immediately append a \DIAOp{FlatMap} operation (line~5) which splits each text line into words and emits one \textcode{std::pair$\langle$std::string,size\_t$\rangle$} (aliased as \textcode{Pair}) containing $(\textcode{word},1)$ per word.
In the example, we use a custom \textcode{Split} function and \textcode{std::string\_view} to reference characters in the text line, and copy them into word strings.
The \textcode{emit} \textcode{auto} parameter of the \DIAOp{FlatMap} lambda function (line~7) enables Thrill to pipeline the \textcode{FlatMap} with the following \DIAOp{ReduceByKey} operation.
Details on pipelining are discussed in Section~\ref{sec:data-flow}.
The result of \DIAOp{FlatMap} is a \DIA{Pair}, which is assigned to the variable \textcode{word\_pairs}.
Note that the keyword \textcode{auto} makes C++ infer the appropriate type for \textcode{word\_pairs} automatically.

The operation \DIAOp{ReduceByKey} is then used to reduce $(\textcode{word},1)$ pairs by \textcode{word}.
This DIA operation must be parameterized with a key extractor (take \textcode{word} out of the pair, line~14) and a reduction function (sum two pairs with the same key together, line~17).
Thrill currently implements \DIAOp{ReduceByKey} using hash tables, as described in Section~\ref{sec:impl details}.
Notice that C++ will infer most types during instantiation of \DIAOp{ReduceByKey}, both input and output are implicit; only with \DIAOp{FlatMap} it is necessary to specify what type gets emitted.

The output of \DIAOp{ReduceByKey} is again a \DIA{Pair}. We need to use a \DIAOp{Map} to transform the \textcode{Pairs} into printable strings (lines~19--21), which can then be written to disk using the \DIAOp{WriteLines} action. Again, the return type of the \DIAOp{Map} (\textcode{std::string}) is inferred automatically, and hence the result of the \DIAOp{Map} operation is implicitly a \DIA{std::string}.

Notice that it is not obvious that the code in Algorithm~\ref{alg:wc} describes a parallel and distributed algorithm.
It is the \emph{implementation} of the DIA operations in the lazily built data-flow graph which perform the actual distributed execution.
The code instructs the C++ compiler to instantiate and optimize these template classes with the UDFs provided.
At runtime, objects of these template classes are procedurally created and evaluated when actions are encountered in the DIA data-flow graph.

\subsection{Overview of DIA Operations}\label{sec:diaop}

\begin{table}
  \caption{DIA Operations of Thrill}
  \label{tab:operations}
  \centering
  \begin{tabular}{|l|l|} \hline
    Operation                                                                   & User Defined Functions                                 \\ \hline\hline
    \multicolumn{2}{|c|}{Sources}                                                                                                        \\ \hline
    $\textbf{Generate}(n) : \Arr{ 0, \ldots, n-1}$                              & $n: \text{DIA size}$                                   \\
    $\textbf{Generate}(n, g) : \Arr{A}$                                         & $g: \textbf{unsigned} \rightarrow A$                   \\ \hline
    $\textbf{ReadLines}() : \text{files} \rightarrow \Arr{\textbf{string}}$     &                                                        \\
    $\textbf{ReadBinary}\langle{A}\rangle() : \text{files} \rightarrow \Arr{A}$ & $A : \text{data type}$                                 \\ \hline\hline
    \multicolumn{2}{|c|}{Local Operations (no communication)}                                                                            \\ \hline
    $\textbf{Map}(f) : \Arr{A} \rightarrow \Arr{B}$                             & $f : A \rightarrow B $                                 \\ \hline
    $\textbf{FlatMap}(f) : \Arr{A} \rightarrow \Arr{B}$                         & $f : A \rightarrow \textbf{list}(B)$                   \\ \hline
    $\textbf{Filter}(f) : \Arr{A} \rightarrow \Arr{A}$                          & $f : A \rightarrow \textbf{bool}$                      \\ \hline
    $\textbf{BernoulliSample}(p) : \Arr{A} \rightarrow \Arr{A}$                 & $p : \text{success probability}$                       \\ \hline
    $\textbf{Union}() : \Arr{A} \times \Arr{A} \cdots \rightarrow \Arr{A}$      &                                                        \\ \hline
    $\textbf{Collapse}() : \Arr{A} \rightarrow \Arr{A}$                         &                                                        \\ \hline
    $\textbf{Cache}() : \Arr{A} \rightarrow \Arr{A}$                            &                                                        \\ \hline\hline
    \multicolumn{2}{|c|}{Distributed Operations (communication between workers)}                                                         \\ \hline
    $\textbf{ReduceByKey}(k,r) : $                                              & $k : A \rightarrow K$                                  \\
    $\textbf{ReduceToIndex}(i,r,n) :$                                           & $i : A \rightarrow [0,n)$                              \\
    \qquad$\Arr{A} \rightarrow \Arr{A}$                                         & $r : A \times A \rightarrow A$                         \\
    $\textbf{GroupByKey}(k,g) :$                                                & $g : \textbf{iterable}(A) \rightarrow B$               \\
    $\textbf{GroupToIndex}(i,g,n) : $                                           & $n : \text{result size}$                               \\
    \qquad$\Arr{A} \rightarrow \Arr{B}$                                         &                                                        \\ \hline
    $\textbf{Sort}(c) : \Arr{A} \rightarrow \Arr{A}$                            & $c : A \times A \rightarrow \textbf{bool} $            \\ \hline
    $\textbf{Merge}(c) : \Arr{A} \times \Arr{A} \cdots \rightarrow \Arr{A}$     & $c : A \times A \rightarrow \textbf{bool} $            \\ \hline
    $\textbf{Concat}() : \Arr{A} \times \Arr{A} \cdots \rightarrow \Arr{A}$     &                                                        \\ \hline
    $\textbf{PrefixSum}(s, i): \Arr{A} \rightarrow \Arr{A}$                     & $s : A \times A \rightarrow A$                         \\
                                                                                & $i : \text{initial value}$                             \\ \hline
    $\textbf{Zip}(z) : \Arr{A} \times \Arr{B} \cdots \rightarrow \Arr{C}$       & $z : A \times B \cdots \rightarrow C $                 \\ \hline
    $\textbf{ZipWithIndex}(z) : \Arr{A} \rightarrow \Arr{B}$                    & $z : \textbf{unsigned} \times A \cdots \rightarrow B $ \\ \hline
    $\textbf{Window}(k, w) : \Arr{A} \rightarrow \Arr{B}$                       & $k : \text{window size}$                               \\
    $\textbf{FlatWindow}(k, f) : \Arr{A} \rightarrow \Arr{B}$                   & $w : A^k \rightarrow B $                               \\
                                                                                & $f : A^k \rightarrow \textbf{list}(B)$                 \\ \hline\hline
    \multicolumn{2}{|c|}{Actions}                                                                                                        \\ \hline
    $\textbf{Execute}()$                                                        &                                                        \\ \hline
    $\textbf{Size}() : \Arr{A} \rightarrow \textbf{unsigned}$                   &                                                        \\ \hline
    $\textbf{AllGather}() : \Arr{A} \rightarrow \textbf{list}(A)$               &                                                        \\ \hline
    $\textbf{Sum}(s, i) : \Arr{A} \rightarrow A$                                & $s : A \times A \rightarrow A$                         \\
    $\textbf{Min}(i) : \Arr{A} \rightarrow A$                                   & $i : \text{initial value}$                             \\
    $\textbf{Max}(i) : \Arr{A} \rightarrow A$                                   &                                                        \\ \hline
    $\textbf{WriteLines}() : \Arr{\textbf{string}} \rightarrow \text{files}$    &                                                        \\
    $\textbf{WriteBinary}() : \Arr{A} \rightarrow \text{files}$                 &                                                        \\ \hline
  \end{tabular}
\end{table}

Table~\ref{tab:operations} gives an overview of the DIA-operations currently supported by Thrill.
The immutability of a DIA enables functional-style data-flow programming.
As DIA operations can depend on other DIAs as inputs, these form a directed acyclic graph (DAG), which is called the \emph{DIA data-flow graph}.
We denote DIA operations as vertices in this graph, and directed edges represent a dependency. Intuitively, one can picture a directed edge as the \emph{values} of a DIA as they flow from one operation into the next.

We classify all DIA operations into four categories. \emph{Source} operations have no incoming edges and generate a DIA from external sources like files, database queries, or simply by generating the integers $0 \ldots n-1$. Operations which have one or more incoming edges and return a DIA are classified further as \emph{local} (LOps) and \emph{distributed} operations (DOps). Examples of LOps are \DIAOp{Map} or \DIAOp{Filter}, which apply a function to every item of the DIA independently. LOps can be performed locally and in parallel, without any communication between workers. On the other hand, DOps such as \DIAOp{ReduceByKey} or \DIAOp{Sort} may require communication and a full data round-trip to disks.

The fourth category are \emph{actions}, which do not return a DIA and hence have no outgoing edges.
The DIA data-flow graph is built lazily, i.e. DIA operations are not immediately executed when created.
Actions trigger evaluation of the graph and return a value to the user program. For example, writing a DIA to disk or calculating the sum of all values are actions.
By inspecting the results of actions, a user program can determine the future program flow, e.g. to iterate a loop until a condition is met.
Hence, control flow decisions are performed collectively in C++ with imperative loops or recursion (host language control-flow).


Initial DIAs can be generated with Thrill's \emph{source} operations. \DIAOp{Generate} creates a DIA by mapping each index $[0,size)$ to an item using a generator function. \DIAOp{ReadLines} and \DIAOp{ReadBinary} read data from the file system and create a DIA with this data.

Thrill's \DIAOp{FlatMap} LOp corresponds to the \emph{map} step in the MapReduce paradigm. Each item of the input DIA is mapped to zero, one, or more output items by a function $f$. In C++ this is done by calling an \DIAOp{emit} function for each item, as shown in the WordCount example. Special cases of \DIAOp{FlatMap} are \DIAOp{Map}, which maps each item to exactly one output, \DIAOp{Filter}, which selects a subset of the input DIA, and \DIAOp{BernoulliSample}, which samples each item independently with constant probability $p$. The LOp \DIAOp{Union} fuses two or more DIAs into one without regard for item order. In contrast, the DOp \DIAOp{Concat} keeps the order of the input DIAs and concatenates them, which requires communication.

\DIAOp{Cache} explicitly materializes the result of a DIA operation for later use. \DIAOp{Collapse} on the other hand folds a pipeline of functions, as described in more detail in Section~\ref{sec:impl details}.

The \emph{reduce} step from the MapReduce paradigm is represented by Thrill's \DIAOp{ReduceByKey} and \DIAOp{GroupByKey} DOps. In both operations, input items are grouped by a key.
Keys are extracted from items using the \emph{key extractor} function $k$, and then mapped to workers using a hash function $h$.
In \DIAOp{ReduceByKey}, the associative reduction function $r$ specifies how two items are combined into one.
In \DIAOp{GroupByKey}, all items with a certain key are collected on one worker and processed by the group function $g$. When possible, \DIAOp{ReduceByKey} should be preferred as it allows local reduction and thus lowers communication volume and running time.

Both ReduceByKey and GroupByKey also offer a \DIAOp{ToIndex} variant, wherein each item of the input DIA is mapped by a function $i$ to an index in the result DIA. The size of the resulting DIA must be given as $n$. Items which map to the same index are either reduced using an associative reduction function $r$, or processed by a group function $g$. Empty slots in the DIA are filled with a neutral item.

\DIAOp{Sort} sorts a DIA with a user-defined comparison function $c$ and \DIAOp{Merge} merges multiple sorted DIAs, again using a user-defined comparison function $c$. \DIAOp{PrefixSum} uses an associative function $s$ to compute the prefix sum (partial sum) for each item.

\DIAOp{Zip} combines two or more DIAs index-wise using a zip function $z$ similar to functional programming languages.
The function $z$ is applied to all items with index $i$ to deliver the new item at index $i$.
The regular \DIAOp{Zip} function requires all DIAs to have equal length, but Thrill also provides variants which cut the DIAs to the shortest or pad them to the longest. \DIAOp{ZipWithIndex} zips each DIA item with its global index. While \DIAOp{ZipWithIndex} can be emulated using \DIAOp{Generate} and \DIAOp{Zip}, the combined variant requires less communication.

\DIAOp{Window} respects the ordering of a DIA and delivers all $k$ consecutive items (a sliding window) to a function $w$ which returns exactly one item. In the \DIAOp{FlatWindow} variant, the window function $f$ can emit zero or more items. Thrill also provides specializations which delivers all disjoint windows of $k$ consecutive items.

\DIAOp{Sum} is an action, which computes an associative function $s$ over all items in a DIA and returns the result on every worker. By default \DIAOp{Sum} uses $+$. \DIAOp{Max} and \DIAOp{Min} are specializations of \DIAOp{Sum} with other operators. \DIAOp{Size} returns the number of items in a DIA and \DIAOp{AllGather} returns a whole DIA as \textcode{std::vector$\langle{T}\rangle$} on each worker. \DIAOp{WriteLines} and \DIAOp{WriteBinary} write a DIA to the file system.  \DIAOp{Execute} can be used to explicitly trigger evaluation of DIA operations.

Besides the actions which trigger evaluation, Thrill also provides \emph{action futures}, \DIAOp{SumFuture}, \DIAOp{MinFuture}, \DIAOp{AllGatherFuture}, etc, which only insert an action vertex into the DIA data-flow graph, but \emph{do not} trigger evaluation.
Using action futures one can calculate multiple results (e.g. the minimum \emph{and} maximum item) with just one data round trip.

The current set of scalable primitive DIA operations listed in Table~\ref{tab:operations} is definitely not final, and more distributed algorithmic primitives may be added in the future as necessary and prudent. In Section~\ref{sec:impl details} we describe the implementations of some of the operations in more detail. We also envision future work on how to accelerate scalable primitives, which can then be use as drop-in replacement to our current straight-forward implementations.

\subsection{Why Arrays?}\label{ss:why}

Thrill's DIA API is obviously similar to Spark and Flink's data-flow languages, which themselves are similar to many functional programming languages~\cite{misale2016comparison}. However, we explicitly define the items in DIAs to be  ordered. This order may be arbitrary after operations like \DIAOp{ReduceByKey}, which hash items to indexes in the array, but they do have an order. Many of our operations like \DIAOp{PrefixSum}, \DIAOp{Sort}, \DIAOp{Merge}, \DIAOp{Zip}, and especially \DIAOp{Window} only make sense with an ordered data type.

Having an order on the distributed array opens up new opportunities in how to exploit this order in algorithms.  Essentially, the order reintroduces the concept of \emph{locality} into distributed data-flow programming.
While one cannot access DIA items directly, such as in a imperative for loop over an array, one \emph{can} iterate over them using a \DIAOp{Window} function \emph{in parallel} with adjacent items as context.
A common design pattern in Thrill programs is to use \DIAOp{Sort} or \DIAOp{ReduceToIndex} to bring items into a desired order, and then to process them using a \DIAOp{Window}.
Furthermore, if the computation in a \DIAOp{Window} needs context from more than one DIA, these can be \DIAOp{Zip}-ped together first.

We are looking forward to future work on how this order paradigm can be exploited.
Furthermore, extending Thrill beyond one-dimensional arrays to higher dimensional arrays, (sparse) matrices, or graphs is not only useful but also conceptually interesting since these data types have a more complex concept of locality.

\subsection{Data-Flow Graph Implementation}\label{sec:data-flow}

Contrary to the picture of DIAs we have drawn for application programmers in the preceding sections, the distributed array of items usually does not exist explicitly.
Instead, a DIA remains purely a conceptual data-flow between two concrete DIA operations.
This data-flow abstraction allows us to apply an optimization called \emph{pipelining} or \emph{chaining}.
Chaining in general describes the process of combining the logic of one or more functions into a single one (called \emph{pipeline}).
In Thrill we chain together all independently parallelizable local operations (\DIAOp{FlatMap}, \DIAOp{Map}, \DIAOp{Filter}, and \DIAOp{BernoulliSample}), and the first local computation steps of the next distributed DIA operation into one block of optimized binary code.
Via this chaining, we reduce both the overhead of the data flow between them, as well as the total number of operations, and obviate the need to store intermediate explicit arrays.
Additionally, we leverage the C++ compiler to combine the local computations \emph{on the assembly level} with full optimization, thus reducing the number of indirections to a minimum, which additionally improves cache efficiency.
In essence, we combine all local computation of one bulk-synchronous parallel (BSP) superstep~\cite{gerbessiotis1994direct} using chaining into one block of assembly code.

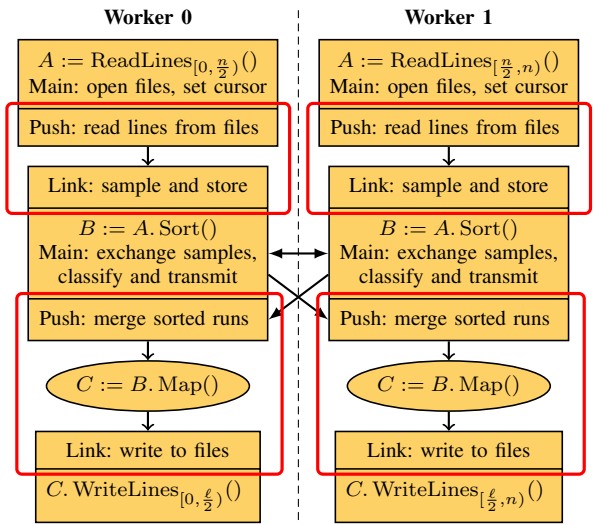
\begin{figure}\centering\footnotesize
  \usetikzlibrary{calc,shapes.multipart}
  \begin{tikzpicture}[
    node distance=2.5mm,
    datanode/.style={draw, semithick, fill=yellow!50!orange!70!white},
    lop/.style={datanode, ellipse, inner sep=0pt, minimum height=19pt},
    dop/.style={datanode, rectangle split, inner sep=4pt, align=center},
    dataarrow/.style={->, thick},
    ]

    \begin{scope}[xshift=0mm]

      \draw[densely dashed] (2.0,1.1) -- ++(0,-6.8);

      \node[dop, rectangle split parts=2] at (0,0) (A0) {
        \nodepart{one} $A := \operatorname{ReadLines}_{[0,\frac{n}{2})}()$ \\
        Main: open files, set cursor
        \nodepart{two} Push: read lines from files
      };

      \node[above=1mm of A0] {\textbf{Worker 0}};

      \node[dop, rectangle split parts=3, below=of A0] (B0) {
        \nodepart{one} Link: sample and store
        \nodepart{two} $B := A.\operatorname{Sort}()$ \\
          Main: exchange samples,\\ classify and transmit
        \nodepart{three} Push: merge sorted runs
      };

      \node[lop,below=of B0] (C0) {$C := B.\operatorname{Map}()$};

      \node[dop, rectangle split parts=2, below=of C0] (F0) {%
        \nodepart{one} Link: write to files
        \nodepart{two} $C.\operatorname{WriteLines}_{[0,\frac{\ell}{2})}()$
      };

      \draw[dataarrow] (A0) -- (B0);
      \draw[dataarrow] (B0) -- (C0);
      \draw[dataarrow] (C0) -- (F0);

    \end{scope}

    \begin{scope}[xshift=40mm]

      \node[dop, rectangle split parts=2] at (0,0) (A1) {
        \nodepart{one} $A := \operatorname{ReadLines}_{[\frac{n}{2},n)}()$ \\
        Main: open files, set cursor
        \nodepart{two} Push: read lines from files
      };

      \node[above=1mm of A1] {\textbf{Worker 1}};

      \node[dop, rectangle split parts=3, below=of A1] (B1) {
        \nodepart{one} Link: sample and store
        \nodepart{two} $B := A.\operatorname{Sort}()$ \\
          Main: exchange samples,\\ classify and transmit
        \nodepart{three} Push: merge sorted runs
      };

      \node[lop,below=of B1] (C1) {$C := B.\operatorname{Map}()$};

      \node[dop, rectangle split parts=2, below=of C1] (F1) {%
        \nodepart{one} Link: write to files
        \nodepart{two} $C.\operatorname{WriteLines}_{[\frac{\ell}{2},n)}()$
      };

      \draw[dataarrow] (A1) -- (B1);
      \draw[dataarrow] (B1) -- (C1);
      \draw[dataarrow] (C1) -- (F1);

    \end{scope}

    \draw[dataarrow, <->, >=latex] (B0.two east) -- (B1.two west);
    \draw[dataarrow, ->, >=latex] ($(B0.two east) + (0,-8pt)$) -- (B1.three west);
    \draw[dataarrow, ->, >=latex] ($(B1.two west) + (0,-8pt)$) -- (B0.three east);

    \begin{scope}[very thick,red,rounded corners=2pt]

      \draw ($(A0.two split west) + (-4pt,-4pt)$) rectangle
      ($(B0.text split east) + (8pt,-2pt)$);

      \draw ($(B0.two split west) + (-4pt,2pt)$) rectangle
      ($(F0.text split east) + (8pt,-2pt)$);

      \draw ($(A1.two split west) + (-4pt,-4pt)$) rectangle
      ($(B1.text split east) + (8pt,-2pt)$);

      \draw ($(B1.two split west) + (-4pt,2pt)$) rectangle
      ($(F1.text split east) + (8pt,-2pt)$);

    \end{scope}

  \end{tikzpicture}
  \caption{Subdivisions of DOps and chained Push, LOps, and Link parts}
  \label{fig:dop parts}
\end{figure}

To integrate the implementations of DIA operations into the pipelining framework we subdivide them into three parts: \emph{Link}, \emph{Main} and \emph{Push} (see Figure~\ref{fig:dop parts} for an example).
The Link part handles incoming items by performing some \emph{finalizing local work} like storing or transmitting them.
This process closes the pipeline and results in a single executable code block containing its logic.
The Main part contains the actual DIA operation logic like sorting, synchronous communication, etc.
And finally, the Push part represents the start of a new pipeline by \emph{emitting} items for further processing.
Depending on the type of a DIA operation, subdivisions can also be empty or trivial.

We explain these subdivisions using \DIAOp{PrefixSum} as an example.
In the Link part, \DIAOp{PrefixSum} receives a stream of items from a preceding operation and stores them in sequence.
While storing them, each worker keeps a local sum over all items.
In the Main part, the workers perform a global synchronous exclusive prefix sum on the local sums to calculate the initial value of their items.
This local initial value is then added to items while they are being read and Push-ed into the next operation.

Chaining also affects how data dependencies between DIA operations are represented in Thrill's data-flow graph.
Due to pipelining of local operations into one assembly block, all LOp are fused with the succeeding DOp vertices.
Hence only vertices representing distributed operations remain in the DAG.
This optimized data-flow DAG corresponds to a set of BSP supersteps and their data dependencies, and is executed lazily when an action is encountered.

Execution is done by Thrill's \emph{StageBuilder}, which performs a reverse breadth-first \emph{stage} search in the optimized DAG to determine which DIA operations need to be calculated.
The gathered vertices are then executed in topological order such that their data dependencies are resolved prior to execution.
Unnecessary recomputations are avoided by maintaining the state of each vertex, and DIA operations are automatically disposed via reference counting.

We implemented chaining and our execution model by making heavy use of C++ template programming.
More precisely, we compose a pipeline by chaining together the underlying (lambda) functions using their static functor types.
Since these types can be deduced by static analysis, chaining can take place during compile time, and hence chained operations can be optimized into single pipelined functions on the assembly code level.
In the end all trivially-parallel local operation like \DIAOp{Map}, \DIAOp{FlatMap}, etc. introduce zero overhead during runtime, and are combined with the following DIA operation's \emph{Link} part.

The caveat of Thrill's chaining mechanism is that the preceding LOp and DOp's (lambda) functions $f_1,f_2,\ldots$ become part of the DIA operation's template instantiation types as \DIA{$T,f_1,f_2,\ldots$}.
This is usually not a problem, since with C++11 we can encourage liberal use of the \textcode{auto} keyword instead of using concrete \DIA{T} types.
However, in iterative or recursive algorithms \DIA{T} variables have to be updated.
These variables are only references to the actual DIA operations, which are immutable, but the references must point to the same underlying DIA operation template type.
We address this issues by adding a special operation named \DIAOp{Collapse} which constructs a \DIA{T} from \DIA{$T,f_1,f_2,\ldots$}.
This operation creates an additional vertex in the data-flow DAG that closes the current pipeline, stores it, and creates a new (empty) one.
The framework will issue compilation errors when \DIAOp{Collapse} is required.

In Thrill we took pipelining of data processing one step further by enabling \emph{consumption} of source DIA storage \emph{while} pushing data to the next operation.
DIA operations transform huge data sets, but a naive implementation would read all items from one DIA, push them all into the pipeline for processing, and then deallocate the data storage.
Assuming the next operation also stores all items, this requires twice the amount of storage.
However, with \emph{consume} enabled, the preceding DIA operation's storage is deallocated while processing the items, hence the storage for all items is needed only once, plus a small overlapping buffer.

\subsection{Data, Network, and I/O Layers}\label{sec:data}

Below the convenient high-level DIA API of Thrill lie several software layers which do the actual data handling.
DIA operations are C++ template classes which are chained together as described in Section~\ref{sec:data-flow}.
These operations store and transmit the items using the \emph{data}, \emph{net}, and \emph{io} layers.

Items have to be serialized to byte-sequences for transmission via the network or for storage on disk. Thrill contains a custom C++ serialization framework which aims to deliver high performance and low to zero overhead. This is possible because neither signatures nor versioning are needed.
In general, fixed-length trivial items like integers and fixed-size numerical vectors are stored with zero overhead. Variable length items like strings and variable-length vectors are prepended with their length. Compound objects are stored as a sequence of their components.

DIA operations process a stream of items, which need to be transmitted or stored, and then read.
Such a stream of items is serialized directly into the memory buffer of a \emph{Block}, which is by default 2\,MiB in size.
Items in a Block are stored without separators or other per-item overhead. This is possible because Thrill's serialization methods correctly advance a cursor to the next item.
Hence, currently only four integers are required as overhead per Block and zero per item.
This efficient Block storage format is important for working with small items like plain integers or characters, but Thrill can also process large blobs spanning multiple Blocks.

A sequence of Blocks is called a \emph{File}, even though it is usually stored in main memory. DIA operations read/write items sequentially to/from Files using template \emph{BlockReader} and \emph{BlockWriter} classes.

To transmit items to other workers, DIA operations have two choices. One is a set of efficient \emph{synchronous} collective communication primitives similar to MPI, such as \emph{AllReduce}, \emph{Broadcast}, and \emph{PrefixSum}. These utilize the same serialization framework and are mostly used for blocking communication of small data items, e.g. an integer AllReduce is often used to calculate the total number of items in a DIA.

The second choice are \emph{Streams} for transmitting large amounts of items \emph{asynchronously}. Streams enable bulk all-to-all communication between all workers.
Thrill contains two subtypes of Streams which differ in the order items are received from other workers: \emph{CatStream}s deliver items strictly in worker rank order, while \emph{MixStream}s deliver items in the arbitrary order in which Blocks are received from the network.
Besides transmitting items in Blocks using the BlockReader and BlockWriter classes, Streams can also scatter whole ranges of a File to other workers without an additional deep copy of the Block's data in the network layer.
Items in Blocks scattered via Streams to workers on the same host are ``transmitted'' via reference counting and not deeply copied.
All communication with workers on the same host is done via shared memory within the same process space.

All Blocks in a Thrill program are managed by the \emph{BlockPool}.
Blocks are reference counted and automatically deleted once they are no longer in any File or used by the network system.
The BlockPool also keeps track of the total amount of memory used in Blocks.
Once a user-defined limit is exceeded, the BlockPool asynchronously swaps out the least recently used Blocks to a local disk.
To distinguish which Blocks may be evicted and which are being used by the data system, Blocks have to be \emph{pinned} to access their data.
Pins can be requested asynchronously to enable prefetching from external memory.
However, all the complexity of pinning Blocks is hidden in the BlockReader/Writer such as to make implementation of DIA operations easy.

Thrill divides available system memory into three parts (by default equally): BlockPool memory, DIA operations memory, and free floating heap memory for user objects like \textcode{std::string}.
All memory is tracked in Thrill by overloading \textcode{malloc()}, hence the user application needs no special allocators.
Memory limits for DIA operations' internal data structures are negotiated and defined when executed during evaluation.
The StageBuilder determines which DIA operations participate in a stage and divides the allotted memory fairly between them.
It is important for external memory support that the operations adhere to these internal memory limits, e.g. by correctly sizing their hash tables and sort buffers.

\subsection{Details on the Reduce, Group, and Sort Implementations}\label{sec:impl details}

Besides pipelining DIA operations, careful implementations of the core algorithms in the operations themselves are important for performance. Most operations are currently implemented rather straight-forwardly, and future work may focus on more sophisticated versions of specific DIA operations. Due to the generic DIA operation interface, these future implementations can then be easily plugged into existing applications.

\subsubsection{Reduce Operations}

\DIAOp{ReduceByKey} and \DIAOp{ReduceToIndex} are implemented using multiple levels of hash tables, because items can be immediately reduced due to associative or even commutative reduction operations $r : A \times A \rightarrow A$.

Thrill distinguishes two reduction phases: the pre-phase prior to transmission and the post-phase receiving items from other workers.
Items which are pushed into the Reduce-DOp are first processed by the key extractor $k : A \rightarrow K$ or index function $i : A \rightarrow [0,n)$ (see Table~\ref{tab:operations}). The key space $K$ or index space $[0,n)$ is divided equally onto the range of workers $[0,p)$. During the pre-phase, each worker hashes and inserts items into one of $p$ separate hash tables, each destined for one worker. If a hash table exceeds its fill-factor, its content is transmitted. If two items with matching keys are found, they are combined locally using $r$.

Items that are received from other workers in the post-phase are inserted into a second level of hash tables. Again, matching items are immediately reduced using $r$.
To enable truly massive data processing, Thrill may spill items into external memory during the post-phase.
The second level of hash tables are again partitioned into $k$ separate tables.
If any of the $k$ tables exceeds its fill-factor, its content is spilled into a File.
When all items have been received by the post-phase, the spilled Files are recursively reduced by choosing a new hash function and reusing the hash table.

The pre- and post-phases use custom linear probing hash tables with built-in reduction on collisions. One large memory segment is used for $p$ separate hash tables. Initially, only a small area of each partition is filled and used to save allocation time. When a hash table is flushed or spilled, its allocated size is doubled until the memory limit prescribed by the StageBuilder is reached.

\subsubsection{Group Operations}

\DIAOp{GroupByKey} and \DIAOp{GroupToIndex} are based on sorting and multiway merging of sorted runs. Items pushed into the Group-DOp are first processed by the key extractor $k : A \rightarrow K$ or index function $i : A \rightarrow [0,n)$, the result space $K$ or $[0,n)$ is distributed evenly onto all $p$ workers. After determining the destination worker, items are immediately transmitted to the appropriate worker via a Stream.
Each worker stores all received items in an in-memory vector.
Once the vector is full or heap memory is exhausted, the vector is sorted by key, and serialized into a File which may be swapped to external memory.
Once all items have been received, the sorted runs are merged using an efficient multiway merger.
The stream of sorted items is separated into subsequences with equal keys, and these sequences are delivered to the group function $g : \textbf{iterable}(A) \rightarrow B$ as a multiway merge iterator.

\subsubsection{Distributed Sorting}

The operation \DIAOp{Sort} rearranges all DIA items into a global order as defined by a comparison function.
In the Link step on each worker, all local incoming items are written to a File.
Simultaneously, a random sample is drawn using reservoir sampling and sent to worker 0 once all items have been seen.
In the Main part, Thrill uses Super Scalar Sample Sort~\cite{sanders2004super} to redistribute items between workers:
worker 0 receives all sample items, sorts them locally, chooses $p-1$ equidistant splitters, and broadcasts the splitters back to all workers.
These build a balanced binary tree with $p$ buckets to determine the target worker for each item in $\lceil \log p \rceil$ comparisons.
As Super Scalar Sample Sort requires the number of buckets to be a power of two, the tree is filled with sentinels as necessary.
Items are then read from the File, classified using the splitter tree, and transmitted via a Stream to the appropriate worker.
When a worker reaches its memory limit while receiving items, the items are sorted and written to a File.
If multiple sorted Files are created, these are merged during the Push part.

Datasets with many duplicated items can lead to load balance problems if sorting is implemented naively.
To mitigate skew, Thrill uses the global array position of the item to break ties and determine its recipient.
When an item is equal to a splitter, it will be sent to the lower rank worker if and only if its global array position is lower than the corresponding quantile of workers.


\section{Experimental Results}\label{sec:experiments}

We compared Apache Spark 2.0.0, Apache Flink 1.0.3, and Thrill using five synthetic micro benchmark applications on the Amazon Web Services (AWS) EC2 cloud.
Our benchmark and input set is based on HiBench~\cite{huang2011hibench}, which we extended\footnote{\url{http://github.com/thrill/fst-bench}} with implementations for Flink and Thrill.

We selected five micro benchmarks: \emph{WordCount}, \emph{\mbox{PageRank}}, \emph{TeraSort}, \emph{KMeans}, and \emph{Sleep}.
To focus on the performance of the frameworks themselves, we attempted to implement the benchmarks equally well using each of the frameworks, and made sure that the same basic algorithms were used.
Spark and Flink can be programmed in Java or Scala, and we include implementations of both whenever possible.
The code for Spark and Flink benchmarks was taken from different sources, all implementations for Thrill were written by us and are included in the Thrill C++ source code as examples.
While we tried to configure Spark and Flink best possible, the complexity and magnitude of configuration options these frameworks provide make it possible that we may have missed some tuning parameters.
For the most part we kept the parameters from HiBench.
The experiments are run with weak scaling of the input, which means that the input size increases with the number hosts $h$, where each AWS host has 32 cores.

\subsection{The Micro Benchmarks}

Implementations of \emph{WordCount} were available in Java and Scala from the examples accompanying Spark and Flink.
The WordCount benchmarks process $h \cdot 32\text{\,GiB}$ of text generated by a C++ version of Hadoop's RandomTextWriter.
There are only 1\,000 distinct words in this random text, which we do not consider a good benchmark for reduce, since only very little data needs to be communicated, but this input seems to be an accepted standard.

For \emph{PageRank} we used only implementations which perform ten iterations of the naive algorithm involving a join of the current ranks with all outgoing edges and a reduction to collect all contributions to the new ranks.
We took the implementation from Spark's examples and modified it to use integers instead of strings as page keys.
We adapted Flink's example to calculate PageRank without normalization and to perform a fixed number of iterations.
Thrill emulates a join operation using \DIAOp{ReduceToIndex} and \DIAOp{Zip} with the page id as the index into the DIA.
The input graph for the experiments contained $h \cdot 4\text{\,M}$ vertices with an average of 39.5 edges per vertex, totaling $\approx h \cdot 2.7\,\text{GiB}$ in size, and generated using the PagerankData generator in HiBench.

\emph{TeraSort} requires sorting 100 byte records, and we used the standard sort method in each framework.
HiBench provided a Java implementation for Spark, and we used an unofficial Scala implementation\footnote{\url{https://github.com/eastcirclek/terasort}}~\cite{dongwon2015terasort} for Flink.
Hadoop's \emph{teragen} was used to generate $h \cdot 16\text{\,GiB}$ as input.

For \emph{KMeans} we used the implementations from Spark and Flink's examples.
Spark calls its machine learning package, while Flink's example is a whole algorithm.
We made sure that both essentially perform ten iterations of Lloyd's algorithm using random initial centroids, and we implemented this algorithm in Thrill.
We fixed the number of dimensions to three, because Flink's implementation required a fixed number of dimensions, and the number of clusters to ten.
Following HiBench's settings, Apache Mahout's GenKMeansDataset was used to generate $h \cdot 16\text{\,M}$ sample points, and the binary Mahout format was converted to text for reading with Flink and Thrill ($\approx h \cdot 8.8\,\text{GiB}$ in size).

The \emph{Sleep} benchmark is used to measure framework startup overhead time.
It launches one map task per core which sleeps for 60 seconds.

\subsection{The Platform}

\begin{table*}[b]
  \caption{Resource Utilization of Frameworks with in Benchmarks}\label{tag:utilization}
  \centering%

  \begin{tabular}{|l|r|r|r|r|r|r|r|r|r|r|r|r|r|} \hline
 & \multicolumn{3}{c|}{Spark (Scala)} & \multicolumn{3}{c|}{Flink (Scala)} & \multicolumn{3}{c|}{Thrill}                                                                               \\ \hline
 & \multicolumn{2}{c|}{CPU}           & \multicolumn{1}{c|}{Net}           & \multicolumn{2}{c|}{CPU} & \multicolumn{1}{c|}{Net} & \multicolumn{2}{c|}{CPU} & \multicolumn{1}{c|}{Net} \\ \hline%

    WordCount & 49\,s  & 64\,\% & 939\,MiB/s & 290\,s & 78\,\%  & 183\,MiB/s & 16\,s & 27\,\% & 1\,127\,MiB/s \\
    PageRank  & 392\,s & 29\,\% & 36\,MiB/s  & 284\,s & 61\,\%  & 151\,MiB/s & 70\,s & 27\,\% & 217\,MiB/s    \\
    TeraSort  & 76\,s  & 20\,\% & 421\,MiB/s & 69\,s  & 26\,\%  & 452\,MiB/s & 49\,s & 25\,\% & 393\,MiB/s    \\
    KMean     & 81\,s  & 27\,\% & 66\,MiB/s  & 183\,s & 4.3\,\% & 29\,MiB/s  & 35\,s & 50\,\% & 253\,MiB/s    \\ \hline
  \end{tabular}

  \medskip
  The table shows the CPU utilization as seconds and percentage of total running time, and the average network bandwidth in MiB/s, both averaged over all hosts during the benchmark run with 16 hosts. TeraSort shows Spark (Java), as we have no Scala implementation.

\end{table*}

We performed our micro benchmarks on AWS using $h$~r3.8xlarge EC2 instances.
Each instance contains 32 vCPU cores of an Intel Xeon E5-2670 v2 with 2.5\,GHz, 244\,GiB RAM, and two local 320\,GiB SSD disks.
We measured 86\,GiB/s single-core/L1-cache, 11.6\,GiB/s single-core/RAM, and 74\,GiB/s 32-core/RAM memory bandwidth using a memory benchmark tool\footnote{\url{http://panthema.net/2013/pmbw/}}.
The SSDs reached 460\,MiB/s when reading 8\,MiB blocks, and 397\,MiB/s when writing.

The $h$ instances were allocated in one AWS availability zone and were connected with a 10 gigabit network.
Our network performance measurements showed $\approx$\,100\,$\mu$s ping latency, and up to 1\,GiB/s sustained point-to-point bandwidth.
All frameworks used TCP sockets for transmitting data.

We experimented with AWS S3, EBS, and EFS as data storage for the benchmark inputs, but ultimately chose to run a separate CephFS cluster on the EC2 instances.
Ceph provided reliable, repeatable performance and minimized external factors in our experiments.
Each EC2 instance carried one Ceph ODS on a local SSD, and we configured the Ceph cluster to keep only one replication block to minimize bandwidth due to data transfer.
We did not use HDFS since Thrill does not support it, and because a POSIX-based distributed file system (DFS) provided a standard view for all frameworks.
The other SSD was used for temporary files created by the frameworks.

All Spark implementations use the RDD interface.
Support for fault tolerance in Spark and Flink incurred no additional overhead, because no checkpoints were written.
By default checkpointing is deactivated and must be explicitly configured.
All compression was deactivated, and Spark was configured to use Kyro serialization.

We used Ubuntu 16.04 LTS (Xenial Xerus) with Linux kernel 4.4.0-31, Ceph 10.2.2 (jewel), Oracle Java 1.8.0\_101, Apache Spark 2.0.0, Apache Flink 1.0.3, and compiled Thrill using gcc 5.4.0 with cmake in Release mode.

\begin{figure*}
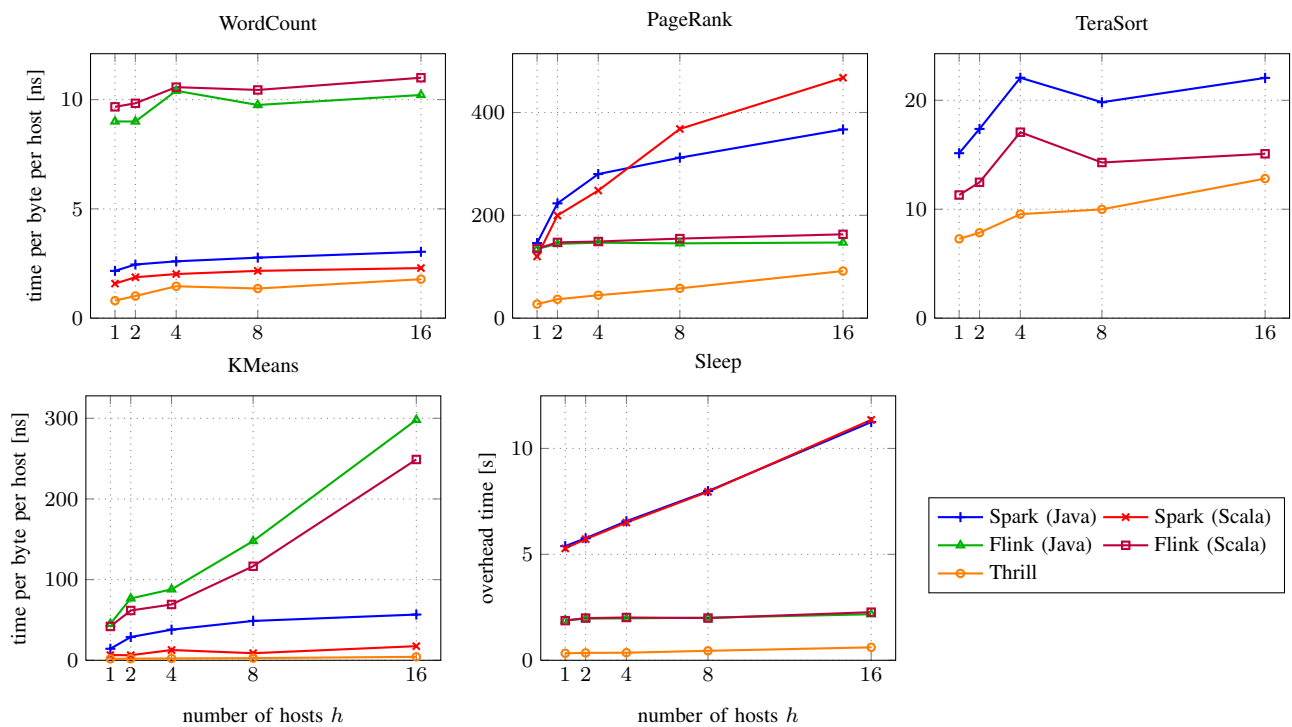

  \centering\footnotesize




  \caption{Experimental results of Apache Spark 2.0.0, Apache Flink 1.0.3, and Thrill on $h$ AWS r3.8xlarge hosts}
  \label{fig:results}
\end{figure*}

\begin{figure*}
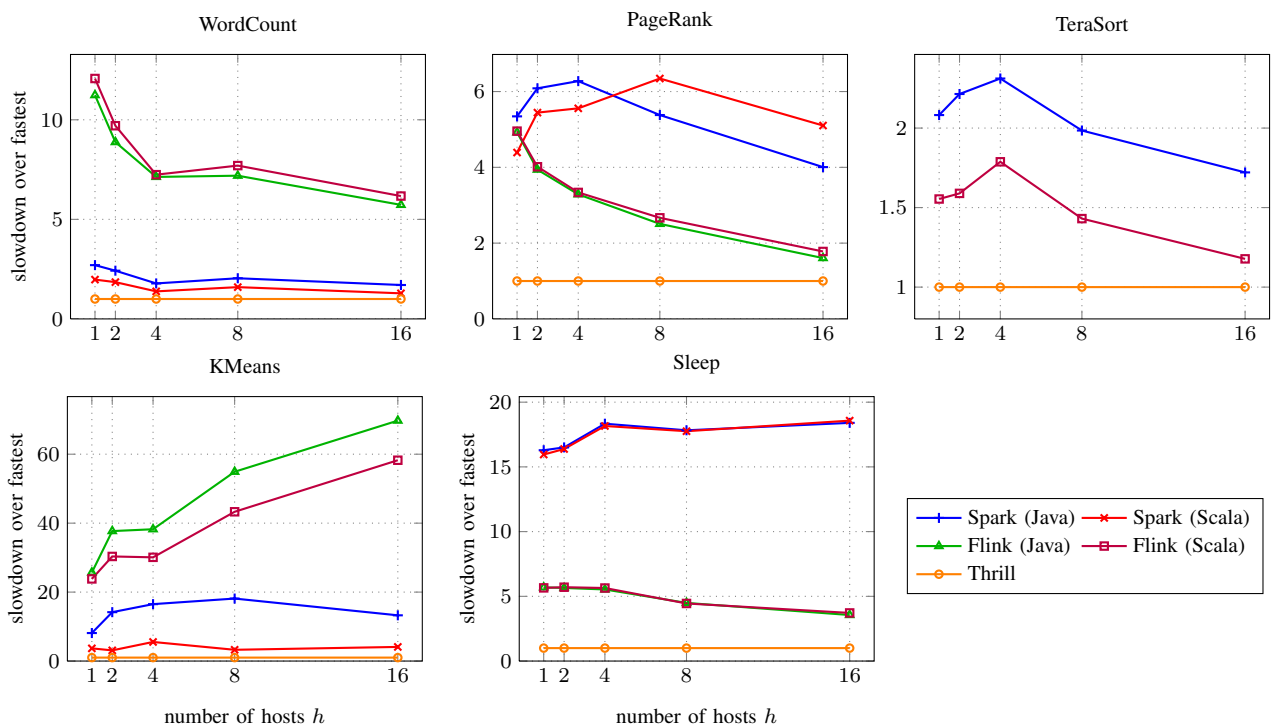

  \centering\footnotesize

  %

  \caption{Slowdown of Apache Spark 2.0.0, Apache Flink 1.0.3, and Thrill on $h$ AWS r3.8xlarge hosts over the fastest framework}
  \label{fig:results slowdown}
\end{figure*}

\def\ProfilePlotSpace{\smallskipamount}
\begin{figure*}
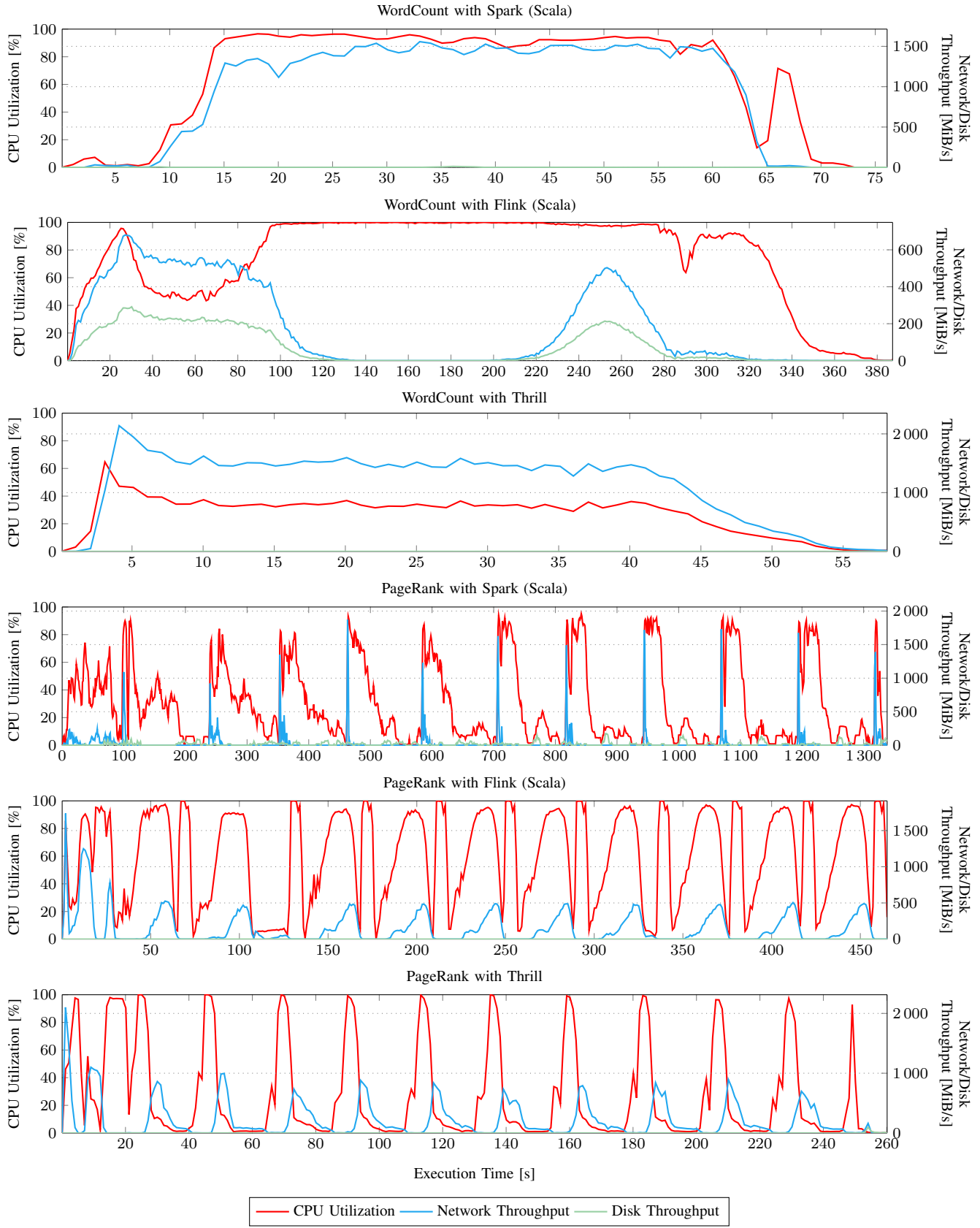

  \centering\footnotesize

  %
%

  \medskip
  \ref{profilelegend}

  \caption{CPU utilization, and network and disk throughput averaged over all hosts during the median WordCount and PageRank benchmark runs with 16 hosts.}
  \label{fig:results profile1}
\end{figure*}

\begin{figure*}
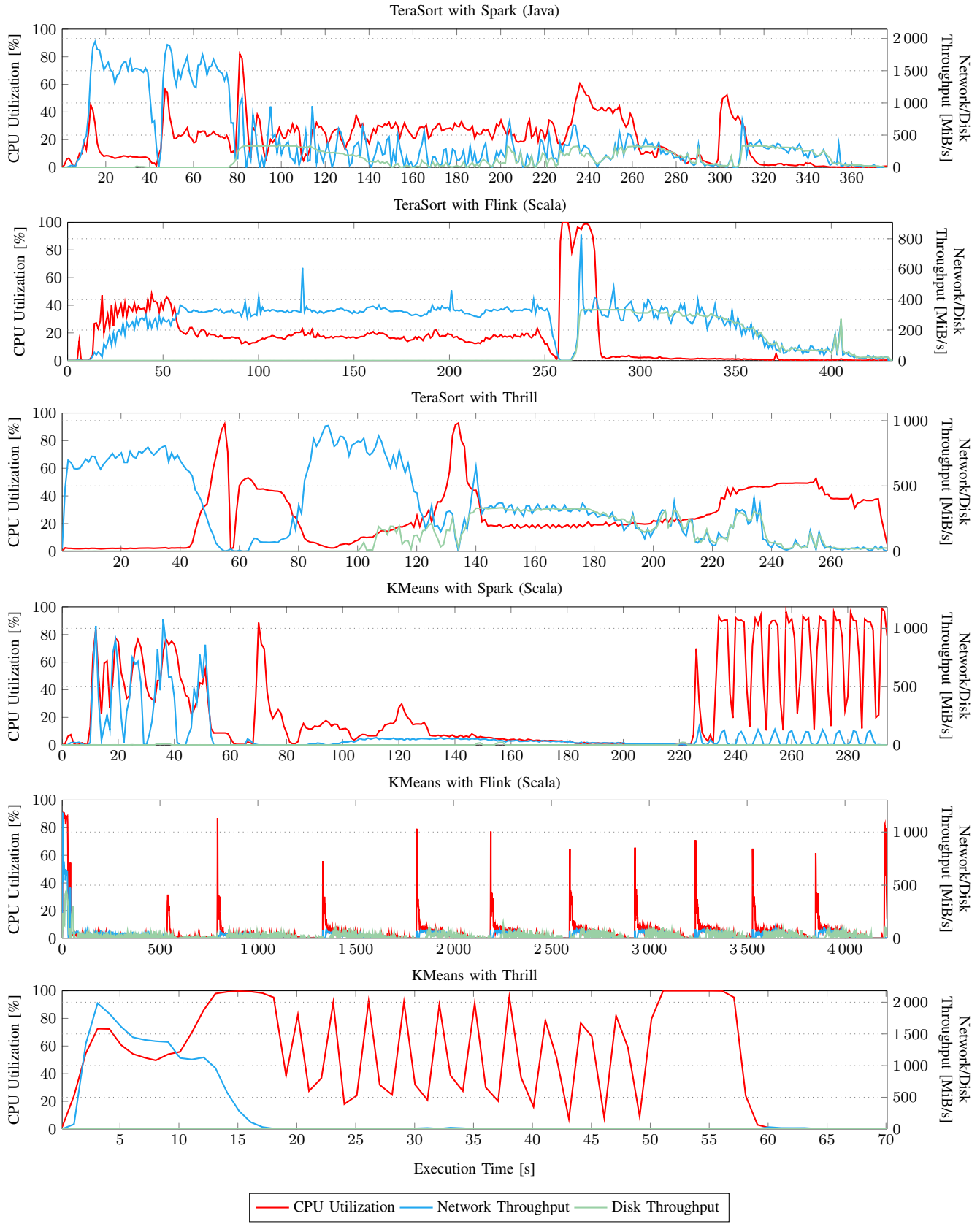

  \centering\footnotesize

  %
%

  \medskip
  \ref{profilelegend}

  \caption{CPU utilization, and network and disk throughput averaged over all hosts during the median TeraSort and KMeans benchmark runs with 16 hosts.}
  \label{fig:results profile2}
\end{figure*}

\subsection{The Results}

Figure~\ref{fig:results} shows the median result of three benchmark runs for $h = 1,2,4,8,16$ hosts.
We plotted the time divided by the number of input bytes on one host, which is proportional to the number of items per host.
Figure~\ref{fig:results slowdown} shows the same results as Figure~\ref{fig:results}, except plotted as the slowdown in running time of each framework over the fastest.
Additionally, we measured a performance profile of the CPU, network, and disk I/O utilization during the benchmarks using information from the Linux kernel, and show the results for $h = 16$ in Table~\ref{tag:utilization} and Figures~\ref{fig:results profile1}--\ref{fig:results profile2}.

Thrill consistently outperforms Spark and Flink in all benchmarks on all numbers of hosts, and is often several times faster than the other frameworks.
The speedup of Thrill over Spark and Flink is highest on a single host, and grows smaller as network and disk I/O become bottlenecks.

In WordCount, the text is read from the DFS, split into words, and the word pairs are reduced locally.
As only 1\,000 unique words occur, the overall result is small and communication thereof is negligible.
Thrill maximizes network utilization with 1\,127\,MiB/s via the DFS and uses only 27\% of the available CPU time for splitting and reducing.
Spark also nearly maximizes the network with 939\,MiB/s, but utilizes the CPU 64\% of the running time.
Flink is a factor 5.7 slower than Thrill in WordCount with 16 hosts, uses the CPU 78\% of the time, and is not network bound.
Thrill's reduction via hash tables are very fast, the other frameworks require considerable more CPU time for the same task.
With 16 hosts Thrill is network bound due to the network file system, and Spark (Scala) is only a factor 1.28 slower.

In PageRank, the current rank values are joined with the adjacency lists of the graph and transmitted via the network to sum all rank contributions for the next iteration in a reduction.
Hence, the PageRank benchmark switches back and forth ten times between high CPU load while joining, and high network load while reducing (see Figure~\ref{fig:results profile1}).
Spark (Java) is a factor 4.0 slower than Thrill on 16 hosts, while Flink (Java) is a factor 1.6 slower.
Flink's pipelined execution engine works well in this benchmark, and reaches 61\% CPU and 15\% network utilization.
From the execution profile of Spark one can see that it does not balance work well between the hosts due to stragglers.
Hence, each iteration takes longer than necessary.
We believe Thrill's performance could be increased even further by implementing a \textcode{Join} algorithm.

In TeraSort, Spark is only a factor 1.7 slower and Flink a factor 1.18 than Thrill on 16 hosts.
Spark reaches only 20\% CPU and 42\% network utilization on average, Flink 26\% and 45\%, and Thrill 25\% and 39\%, respectively.
Flink's pipelined execution outperforms Spark in TeraSort, as was previously shown by an other author~\cite{dongwon2015terasort}.
The implementations appear well tuned, however, due to the CPU and network utilization we believe all can be improved.

In the KMeans algorithm, the set of centroids are broadcast.
Then all points are reclassified to the closest centroid, after which new centroids are determined from all points via a reduction.
Like PageRank, the KMeans algorithm interleaves high local work and high network load (a reduction and a broadcast).
Spark (Scala) is a factor 4.1 slower than Thrill on 16 hosts, Spark (Java) a factor 13, and Flink more than 50.
We believe this is due to the JVM object overhead for vectors, and to inefficiencies in the way Spark and Flink broadcast the centroids.
Flink's query optimizer does not seem to work well for the KMeans example accompanying their source package.
Thrill utilizes the CPU 50\% and the network 25\% of the running time, while Spark reach 27\% CPU and only 7\% network utilization.

The Sleep benchmark highlights the startup time of the frameworks.
We plotted the running time excluding the slept time in Figure~\ref{fig:results}.
Spark requires remarkably close to $5 + h \cdot 0.4$ seconds to start up.
Apparently, hosts are not started in parallel.
Flink's start up time was much lower, and Thrill's less than one second.


\section{Conclusion and Future Work}\label{sec:conclusion}

With Thrill we have demonstrated that a C++ library can be used as a distributed data processing framework reaching a similarly high level of abstraction as the currently most popular systems based on Java and Scala while gaining considerable performance advantages.
In the future, we want to use Thrill on the one hand for implementing scalable parallel algorithms (e.g. for construction of succinct text indices) that are both advanced and high level.
Thrill has already been used for more than five suffix sorting algorithms, logistic regression, and graph generators.
On the other hand, at a much lower level, we want to use Thrill as a platform for developing algorithmic primitives for big data tools that enable massively scalable load balancing, communication efficiency, and fault tolerance.

While Thrill is so far a prototype and research platform, the results of this paper are sufficiently encouraging to see a possible development into a main stream big data processing tool.
Of course, a lot of work remains in that direction such as implementing interfaces for other popular tools like Hadoop and the AWS stack, and creating frontends in scripting languages like Python for faster algorithm prototyping.
To achieve practical scalability and robustness for large cluster configurations, we also need significant improvements in issues like load balancing, fault tolerance and native support for high performance networks like InfiniBand or Omni-Path.

Furthermore, we view it as useful to introduce additional operations and data types like graphs and multidimensional arrays in Thrill (see also Section~\ref{ss:why}).
But, we are not sure whether automatic query plan optimization as in Flink should be a focus of Thrill, because that makes it more difficult to implement complex algorithms with a sufficient amount of control over the computation.
Rather it may be better to use Thrill as an intermediate language for a yet higher level tool that would no longer be a plain library but a true compiler with a query optimizer.

\section*{Acknowledgment}
We would like thank the AWS Cloud Credits for Research program for making the experiments in Section~\ref{sec:experiments} possible.
Our research was supported by the Gottfried Wilhelm Leibniz Prize 2012, and the Large-Scale Data Management and Analysis (LSDMA) project in the Helmholtz Association.

\IEEEtriggeratref{18}


\bibliographystyle{IEEEtran}
\bibliography{IEEEabrv,references}





\end{document}